# Quantum Correlation of Microwave Two-mode Squeezed State Generated by Nonlinearity of InP HEMT

A. Salmanogli

Cankaya University, Engineering Faculty, Electrical and Electronic Department, Ankara, Turkey

**Abstract**

This study significantly concentrates on cryogenic InP HEMT high-frequency circuit analysis using quantum theory to find how the transistor nonlinearity can affect the quantum correlation of the modes generated in the circuit. Firstly, the total Hamiltonian of the circuit is derived, and the dynamic equation of the motion contributed is examined using the Heisenberg-Langevin equation. Using the nonlinear Hamiltonian, some components are attached to the intrinsic internal circuit of InP HEMT to fully address the circuit characteristics. The components attached are arisen due to the nonlinearity effects. As a result, the theoretical calculations show that the states generated in the circuit are mixed, and no pure state is produced. Accordingly, the modified circuit generates the two-mode squeezed thermal state, which means one can focus on calculating the Gaussian quantum discord to evaluate quantum correlation. It is also found that the nonlinearity factors (addressed as the nonlinear components in the circuit) can intensely influence the squeezed thermal state by which the quantum discord is changed. Finally, as the primary point, it is concluded that although it is possible to enhance the quantum correlation between modes by engineering the nonlinear components; however, quantum discord greater than unity, entangled microwave photons, seems a challenging task since InP HEMT operates at 4.2 K.

**Introduction**

Quantum correlation as a fundamental issue is significantly represented in quantum applications such as quantum information [1-7] and quantum sensors [8-12]. The entanglement has been synonymously applied in the same way with quantum correlation in most applications since the quantum system contains only pure states [5]. In contrast, when the mixed states are generated by a quantum system, which covers most quantum systems, such as quantum radars [8, 10], the term entanglement cannot be used over quantum correlation. It is because some separable mixed states can introduce "residual correlation" that cannot be fitted by any classical probability distributions [5-7]. In other words, for a bipartite system with separable state $\rho_{AB}$ expressed as $\rho_{AB} = \sum P_i\, \rho_{Ai} \times \rho_{Bi}$, where $\rho_{Ai}$ and $\rho_{Bi}$ are density matrices of the subsystems; the states $\rho_{Ai}$ and $\rho_{Bi}$ may be physically non-distinguishable. Consequently, all information about the subsystems cannot be locally retrieved due to the nonorthogonality of the states [7]. Some recently published studies have shown that classically correlated states might show the signature of quantumness [13-14]. To evaluate the quantumness, the term "quantum discord" has been applied [1-7]. The quantum discord can quantify the residual correlation or the signature of quantumness [5-7], which captures all quantum correlations in a bipartite state. However, several different quantifiers have been introduced to evaluate the nonclassicality of the correlation (quantum correlation); but the most popular one is quantum discord [15]. Quantum discord is the difference between the quantum correlation within a quantum state and its classical correlations [5-7]. Some studies show that quantum discord can be defied for both qubits and continuous variable systems [7, 16]. Therefore, studying the quantum discord for the continuous variables is valuable due to the critical applications of the continuous variables, such as in quantum computation and quantum communications [18]. For some reason mentioned above, and since the circuit designed in this work interacts with the environment as any real quantum systems interact inevitably with the surrounding, we have to focus on the continuous variable and calculate the quantum discord for these variables. As the main point, the influence of the thermal noise generated by the circuit is studied on the quantum correlation. Thus, concerning the mentioned points, the study significantly focuses on the "Gaussian quantum discord" and uses the close formula satisfied to all families of the Gaussian state. Accordingly, the Gaussian state includes the important class of the squeezed thermal state. This unique state is realized by applying two-mode squeezing to a pair of single-mode thermal states [5, 7]. Recent studies have shown that the squeezed thermal state, generally the Gaussian state, can be decomposed into the EPR (Einstien-Podolsky-Rosen) state plus the local action of a "phase-sensitive Gaussian channel" [5].

As mentioned, this study significantly focuses on the two-mode squeezing thermally state, which is generated by the nonlinearity of the InP HEMT. It has to be initially shown that due to the nonlinearity of InP HEMT in the circuit, the states generated in the system are mixed. This means that the quantum discord as a quantifier can completely define the quantum correlation rather than the entanglement. Then, it will be verified that the introduced nonlinearity in the circuit generates the microwave two-mode squeezed thermal state. Finally, it introduces some critical factors (components) relating to nonlinearity to enhance the quantum correlation between modes. The mentioned components, as the circuit elements, are attached to the original internal circuit of InP HEMT to make the modified circuit; Then, the modified circuit is simulated using PySpice. The simulation in PySpice is performed to check some critical issues from a circuit analysis point of view. Consequently, it will show that the attached nonlinearity can add a challenging trade-off.

It is better to indicate that this study completes the latter similar work [17], in which, for simplicity, it was supposed that all states produced by the InP HEMT are pure states. Another difference between the present work with [17] is that this study concentrates on the microwave two-mode squeezed thermal state and the squeezing parameters that affect the quantum correlation. Additionally, in this study, the thermally excited photons in the Drain due to the related conductance are introduced at 450 K (Drain conductance noise temperature), making the circuit behave like an accurate model.

The present study initially defines the system and its crucial elements that can affect the primary goal. In the next step, the Hamiltonian of the system is derived, and using the nonlinear part of the Hamiltonian, the dynamics equation of the motion is introduced. In addition, it is shown that the circuit can generate a microwave two-mode squeezed thermal state by applying some nonlinear coefficients. In the other step, it is theoretically shown that the introduced circuit generates the mixed states. Finally, the theory related to quantum discord and its deriving is introduced and discusses by which parameters or factors one can manipulate the quantum discord.

## Theory and Background

*System definition*

The circuit studied in this work is schematically shown in Fig. 1. It contains InP HEMT as a nonlinear active element, an input and output matching network to match the input and output impedance, and a DC stabilization circuit. The generic equivalent circuits of the input and output matching network are shown in the inset figures. In the inset figures, it is shown in a usual way that any transmission line can be modeled with an equivalent lumped element circuit. As shown in the schematic, InP HEMT is biased via $V_g$ and $V_d$ to operate in the desired region, and the small signal RF input wave is applied to the circuit through an input capacitor ($C_{in}$). $L_{g1}$ and $L_{d1}$ are key elements in the stabilization circuit. As an essential element, InP HEMT's nonlinear equivalent circuit is attached as another inset figure. The attached inset tries to completely show all factors that could affect the operation of InP HEMT at cryogenic temperature; it is mainly for quantum applications in which the number of microwave photons is dramatically kept at a low level. For the reason that the thermally excited photons can affect the low photons quantum applications, the noise generated by resistances in the circuit is considered. This makes the circuit behave like a real one. For instance, $4KTB_nR_s$ is the voltage-noise generated by $R_s$ in the circuit, in which K, T, and $B_n$ are, respectively, Boltzmann constant, operational temperature, and noise bandwidth.

One of the essential components in the inset circuit is $i_{ds}$, defined as the dependent current source controlled by the voltage. As can be seen from the related relationship, the amount of the current is controlled by $g_m$, defined as the intrinsic transconductance of the circuit; that is additionally manipulated by $g_{m2}$ and $g_{m3}$ called the second and third nonlinearity factors (generally called nonlinearity factors). This paper focuses on the latter factors and their effect on the circuit, through which the circuit's resonance frequency is changed. In addition, the coupling signals between the resonators in the circuit can be strongly changed. The term "coupling between the resonators" is used because, from a general view, the circuit illustrated in Fig. 1 can be supposed as the two separate oscillators oscillating in the gate and drain side of the transistor, which are coupled to each other through the InP HEMT nonlinear circuit. In other words, the nonlinearity in InP HEMT changes the features of the oscillators, including the resonance frequency and their impedance.

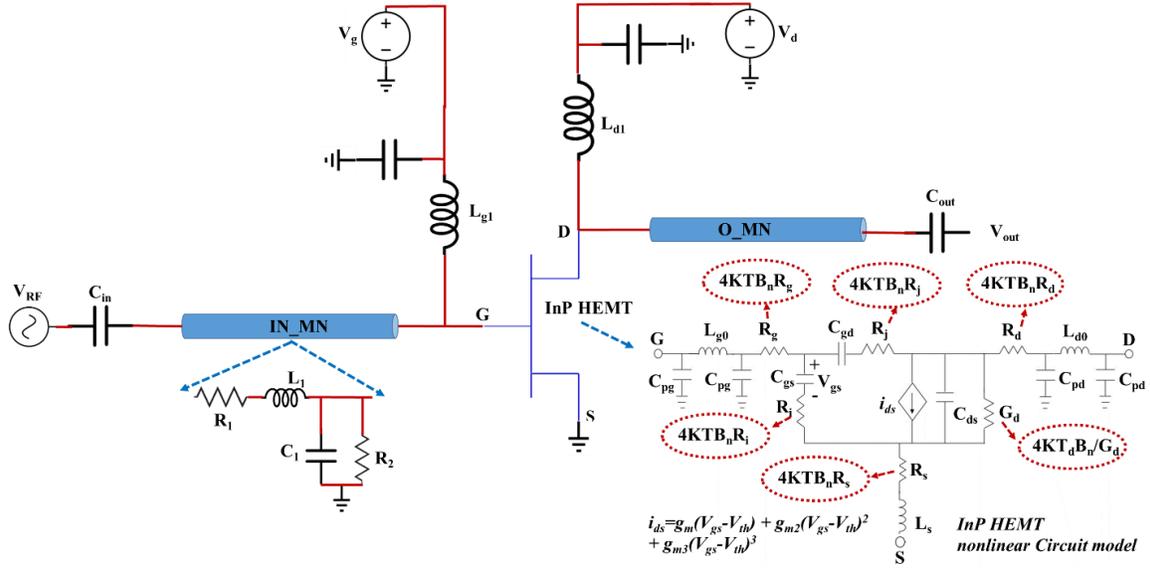

Fig. 1 the schematic of the circuit containing the input and output matching networks, stability network, and active elements (InP HEMT) and its internal circuit.

*Hamiltonian of the circuit*

To derive the classical Hamiltonian, one has to use Legendre transformation as $H(\varphi_k, Q_k) = \sum_k (\dot{\varphi}_k \cdot Q_k) - L_c$, where $L_c$ and $Q_k$ are, respectively, the system's Lagrangian and conjugate variables of the "coordinate variable" $\varphi_k$, calculated through $Q_k = \partial L_c / \partial (\partial \varphi_k / \partial t)$ [17, 20]. The total Hamiltonian of the circuit introduced in Fig. 1 is given by:

$$H_t = \frac{C_A}{2}\dot{\varphi}_1^2 + \frac{1}{2L_1}\varphi_1^2 + \frac{C_B}{2}\dot{\varphi}_2^2 + \frac{1}{2L_2}\varphi_2^2 - C_c\dot{\varphi}_1\dot{\varphi}_2 + \left\{g_{m2}\varphi_2\dot{\varphi}_1^2 + 2g_{m3}\varphi_2\dot{\varphi}_1^3\right\}$$
$$-\varphi_1\left(\overline{I_g^2} - \overline{I_i^2}\right) - \varphi_2\left(\overline{I_{ds}^2} + \overline{I_d^2} + \overline{I_j^2}\right) - \frac{C_{in}}{2}V_{rf}^2 - \frac{C_{gs}}{2}\overline{V_i^2}$$

(1)

where $C_A = C_{in} + C_1 + C_{gs} + C_{gd}$, $C_B = C_{gd} + C_2$, and $C_c = C_{gd}$. The total Hamiltonian of the system is written as $H_t = H_L + H_N$, where $H_L$ stands for linear Hamiltonian and $H_N$ contains the nonlinear terms. The nonlinearity is arisen due to the terms inside the curly bracket. This study looks to generate two-mode squeezed thermal states using the nonlinearity introduced by InP HEMT. For this reason, the focus is just laid on $H_N$ and its analysis, using which I. it will theoretically show that the generated states by the circuit presented in Fig. 1 are mixed, and II. there is the nonlinearity effect that can generate the two-mode squeezed thermal state. In contrast, $H_L$ is used to define the steady-state operational point, energy level, and also the linear part of Hamiltonian has some terms that can generate the coherent state. The definition of

$H_L$ and its parameters have been introduced in Appendix A (Eq. A1 and A2). Consequently, the nonlinear Hamiltonian is given by:

$$H_N = \frac{g_{N2}}{C_M^4}\{C_B^2\varphi_2 Q_1^2 + C_c^2\varphi_2 Q_2^2 + g_m^2 C_B^2\varphi_2^3 - 2g_m C_B^2 Q_1\varphi_2^2 - 2g_m C_B C_c Q_2\varphi_2^2 + 2g_m C_B C_c Q_1\varphi_1\varphi_2\} \quad (2)$$

where $g_{N2} = g_{m2} + 6g_{m3}[\partial\varphi_1/\partial t]_{DC}$, $C_M^2 = C_B(C_A+C_N)-C_c^2$, and $C_N = g_{m2}[\varphi_2]_{DC} + 6g_{m3}[\varphi_2]_{DC}*[\partial\varphi_1/\partial t]_{DC}$. In the defined relationships, $g_{N2}$ and $C_N$ are estimated using the approximation methods and are, respectively, defined as the nonlinearity factor and nonlinear capacitance. It is clearly shown in Eq. 2 that the $g_{N2}$ and $C_N$ strongly manipulate the amplitude of the nonlinearity created in InP HEMT. Additionally, it is shown that the resonance frequency of the second oscillator is dramatically dependent on $C_N$. This means that the nonlinearity in InP HEMT can affect the resonance frequency, which is calculated as $\omega_2 = 1/\sqrt{(C_{q2}L_{2'})}$; in this equation, $C_{q2}$ and $L_{2'}$ expressed in detail in Appendix A (Eq. A2) are affected by the nonlinearity factors. To clarify the effects of the nonlinearity, some components created because of the InP HEMT nonlinearity are attached to the main circuit (inset figure in Fig. 1) and illustrated in Fig. 2. These elements are theoretically derived using nonlinear Hamiltonian expressed in Eq. 2. The red scribble-dotted line on the circuit contains a variable capacitor ($C_N$) and controllable current depending on the nonlinearity factors of the InP HEMT.

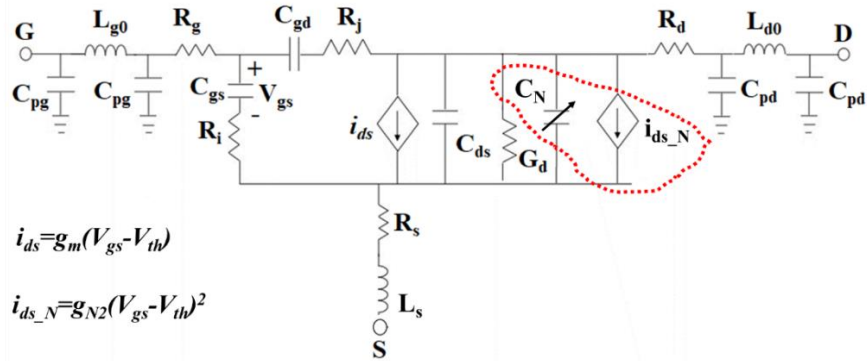

Fig.2 An approximation of the InP HEMT internal circuit by considering the nonlinearity effects illustration with a scribble-dotted line on the primary circuit.

The circuit illustrated in Fig. 2 is an approximated circuit that may model the InP HEMT transistor behavior. The present study uses this model to quantum mechanically analyze the InP HEMT nonlinearity effect on the quantum correlation that may occur between modes generated in the circuit at cryogenic temperature. As clearly seen in Fig. 2, $C_N$ and $i_{ds\_N}$ directly affect the second oscillator resonance properties; nonetheless, the nonlinear effects are coupled to the first oscillator through $C_{gd}$. This means that the coupling properties of the oscillators are strongly affected by the nonlinear properties. For example, the cross-correlation between the coupled oscillator's modes is severely influenced by the nonlinearity factors defined in the circuit. In other words, the classicality correlated modes generated by the coupled oscillators can be affected in such a way as to show quantumness. This phenomenon strongly depends on nonlinearity and its impacts.

As an important quantifier, the quantum discord [5-7] is selected to evaluate the quantum correlation between the states generated in the circuit. The quantum discord rather than the entanglement is chosen to show the effects because the circuit discussed in this study shows mixed states. This point will be explored in the next section. In the complementary part of this study, the circuit shown in Fig. 2 is analyzed using PySpice to investigate the effect of the attached components on InP HEMT's main feature, such as DC characterization. Using PySpice, rather than any specific CAD simulator tools, gives some degree of freedom by which the designer can easily attach any elements or sub-elements to the circuit and make the desired pack. We also selected PySpice to work with because all theoretical simulations to calculate the mean photons number, quantum discord, the smaller Symplectic eigenvalue, and quantum mutual information were done in Python.

It should be noted that the general aim of the attached elements to the circuit is to study their effects on the performance of InP HEMT operating at 4.2 K and determine by what factors it can manipulate the quantum correlation between modes generated in the circuit. In the following, it is necessary to re-express Eq. 2 in the form of the ladder operator to analyze the quantum correlation between the modes generated by the circuit's nonlinearity. Using the traditional methods, the nonlinear Hamiltonian in terms of the ladder operator is expressed as:

$$H_N = \frac{g_{N2}}{C_M^4}\left\{-C_B^2\sqrt{\frac{\hbar Z_2}{2}}\frac{\hbar}{2Z_1}(a_2+a_2^+)(a_1-a_1^+)^2 - C_c^2\sqrt{\frac{\hbar Z_2}{2}}\frac{\hbar}{2Z_2}(a_2+a_2^+)(a_2-a_2^+)^2 \right.$$
$$+ g_m^2 C_B^2\sqrt{\frac{\hbar Z_2}{2}}\frac{\hbar}{2Z_2}(a_2+a_2^+)^3 + i2g_m C_B^2\sqrt{\frac{\hbar}{2Z_1}}\frac{\hbar Z_2}{2}(a_1-a_1^+)(a_2+a_2^+)^2 \qquad (3)$$
$$\left. + i2g_m C_B C_c\sqrt{\frac{\hbar}{2Z_2}}\frac{\hbar Z_2}{2}(a_2-a_2^+)(a_2+a_2^+)^2 - i2g_m C_B C_c\sqrt{\frac{\hbar}{2Z_1}}\frac{\hbar\sqrt{Z_1 Z_2}}{2}(a_1-a_1^+)(a_1+a_1^+)(a_2+a_2^+)\right\}$$

where $a_k$ and $a_k^+$ (k = 1, 2) are the annihilation and creation operators, respectively. For simplicity, a few constants are defined by which Eq. 3 is simplified as:

$$H_N = \left\{-\hbar g_{N11}(a_2+a_2^+)(a_1-a_1^+)^2 - \hbar g_{N21}(a_2+a_2^+)(a_2-a_2^+)^2 + \hbar g_{N31}(a_2+a_2^+)^3 \right.$$
$$\left. + i\hbar g_{N41}(a_1-a_1^+)(a_2+a_2^+)^2 + i\hbar g_{N51}(a_2-a_2^+)(a_2+a_2^+)^2 - i\hbar g_{N61}(a_1-a_1^+)(a_1+a_1^+)(a_2+a_2^+)\right\} \qquad (4)$$

where $g_{N11}$, $g_{N21}$, $g_{N31}$, $g_{N41}$, $g_{N51}$, and $g_{N61}$, are the related constants defined in Appendix A (Eq. A3). These constants are strongly dependent on the nonlinearity factor. The dynamics equation of motion of the circuit is calculated using the Heisenberg-Langevin equation; A simple way to achieve a stationary and robust calculation in continuous modes to calculate quantum correlation, i.e. entanglement [28-29] or quantum discord [9], is to select a constant point that the system is driven and works with. Since the interaction field is so strong, it is appropriate to focus on linearization and calculate the quantum fluctuation around the semi-classical constant point. For linearization, the oscillator modes are expressed as the summation of the stationary (constant) and fluctuating parts as $\mathbf{a_1} = A_1+\boldsymbol{\delta a_1}$ and $\mathbf{a_2} = A_2+\boldsymbol{\delta a_2}$, where the capital letter ($A_1$ and $A_2$) denotes the system's steady-state points, and $\boldsymbol{\delta}$ indicates the fluctuation around the steady-state point. The steady-state points related to the circuit are calculated and presented in Appendix A (Eq. A7). Thus, the linearized equations around the steady-state points are given by:

$$\dot{\delta a_1} = -\left(i\omega_1 + \frac{\kappa_1}{2}\right)\delta a_1 + \gamma_{a11}\left(\delta a_1 + \delta a_1^+\right) + \gamma_{a12}\left(\delta a_2 + \delta a_2^+\right) + \gamma_{a13}\delta a_1^+ + \sqrt{2\kappa_1}\delta a_{in-1}$$
$$\dot{\delta a_2} = -\left(i\omega_2 + \frac{\kappa_2}{2}\right)\delta a_2 + \gamma_{a21}\left(\delta a_1 - \delta a_1^+\right) + \gamma_{a22}\left(\delta a_2 + \delta a_2^+\right) + \gamma_{a23}\left(\delta a_2 - \delta a_2^+\right) + \gamma_{a24}\left(\delta a_1 + \delta a_1^+\right) + \sqrt{2\kappa_2}\delta a_{in-2}$$
(5)

where $\gamma_{a11}$, $\gamma_{a12}$, $\gamma_{a13}$, $\gamma_{a21}$, $\gamma_{a22}$, $\gamma_{a23}$, and $\gamma_{a24}$ are constant rates depending on the stationary points of the system. These rates can be complex numbers defined in Appendix A (Eq. A4). In Eq. 5, $\kappa_1$, $\kappa_2$, $\omega_1$, and $\omega_2$ are the first and second oscillators' decay rates and the contributed frequencies, respectively. Additionally, $\delta a_{in\_1}$ and $\delta a_{in\_2}$ are the input noises fluctuation. They obey the correlation function $< \delta a_{in\_1}(s)\delta a_{in\_1}^+(s') > = [1+ N(\omega)]\times\boldsymbol{\delta}(s-s')$, where $N(\omega) = [\exp(\hbar\omega/k_BT)-1]^{-1}$, in which $k_B$ and T are the Boltzmann's constant and operational temperature, respectively [9, 17, 20]. $N(\omega)$ is the equilibrium mean of the thermal photon numbers at the frequency $\omega$. Finally, capital $\boldsymbol{\delta}$ expressed in $\boldsymbol{\delta}$(s-s') is Dirac's function. In the following, Eq. 5 is transformed to the Fourier domain to simplify the algebra in the frequency domain to calculate the quantum discord, which is introduced as:

$$\left[i\Delta_1 + \frac{\kappa_1}{2} - \gamma_{a11}\right]\delta a_1 = \left(\gamma_{a11} + \gamma_{a13}\right)\delta a_1^+ + \gamma_{a12}\left(\delta a_2 + \delta a_2^+\right) + \sqrt{2\kappa_1}\delta a_{in-1}$$
$$\left[i\Delta_2 + \frac{\kappa_2}{2} - \gamma_{a22} - \gamma_{a23}\right]\delta a_2 = \gamma_{a21}\left(\delta a_1 - \delta a_1^+\right) + \left(\gamma_{a22} + \gamma_{a23}\right)\delta a_2^+ + \gamma_{a24}\left(\delta a_1 + \delta a_1^+\right) + \sqrt{2\kappa_2}\delta a_{in-2}$$
(6)

where $\Delta_1$ and $\Delta_2$ are the oscillators detuning frequencies, which are calculated as $\Delta_1 = \omega - \omega_1$ and $\Delta_2 = \omega - \omega_2$, where $\omega$ is the RF incident frequency. Notably, Eq. 6 is used to calculate the $< \delta a_1^+\delta a_1>$, $< \delta a_2^+\delta a_2>$, and $< \delta a_1\delta a_2>$ as the mean photon number of the first-, second-oscillator, and the phase-sensitive cross-correlation [17-20].

*Mixed states generation because of the nonlinearity of InP HEMT*

This part shows that the state of the oscillators is dispersed, meaning that all states generated by the circuit become mixed states. One can utilize the first perturbation theory to calculate the oscillator's state changes. The change of a typical state is calculated using the first perturbation theory by $|j>^{(1)} = \sum_{i\neq j} \{(<i|H_N|j>)/(E_i-E_j)\}\times|j>$ [20], where $|j>$ is the pure state of the oscillators, and $E_i$ and $E_j$ are the contributed energies. $|j>^{(1)}$ is the final state of the first oscillator that may differ from $|j>$; it depends on the system and the related Hamiltonian. The results of the calculation are presented in Eq. 7 as:

$$|j_1\rangle^{(1)} = \sum_{i\neq j_1} \frac{\langle i|H_N|j_1\rangle}{E_i - E_{j1}}|j_1\rangle = \frac{j_{p11}}{E_{(j_1-2)} - E_{j1}}|j_1-2\rangle + \frac{j_{p12}}{E_{(j_1+2)} - E_{j1}}|j_1+2\rangle + \frac{j_{p13}}{E_{(j_1-1)} - E_{j1}}|j_1-1\rangle - \frac{j_{p13}}{E_{(j_1+1)} - E_{j1}}|j_1+1\rangle$$

$$|j_2\rangle^{(1)} = \sum_{i\neq j_1} \frac{\langle i|H_N|j_2\rangle}{E_i - E_{j2}}|j_2\rangle = \frac{j_{p21}}{E_{(j_2-3)} - E_{j2}}|j_2-3\rangle + \frac{j_{p21}}{E_{(j_2+3)} - E_{j2}}|j_2+3\rangle + \frac{[j_{p23} + j_{p27} + i(j_{p26} + j_{p27} + j_{p28})]}{E_{(j_2-1)} - E_{j2}}|j_2-1\rangle$$

$$+ \frac{[j_{p24} + j_{p25} + i(j_{p25} + j_{p29} + j_{p210})]}{E_{(j_2+1)} - E_{j2}}|j_2+1\rangle + \frac{j_{p211}}{E_{(j_2+2)} - E_{j2}}|j_2+2\rangle + \frac{j_{p212}}{E_{(j_2-2)} - E_{j2}}|j_2-2\rangle$$

(7)

In this equation, the data related to the constants used are given in Appendix A (A6). This equation shows that the final states related to the first and second oscillators $|j_1\rangle^{(1)}$ and $|j_2\rangle^{(1)}$, are strongly influenced by the nonlinear (perturbation) Hamiltonian. For example, the state of the first oscillator is coupled to $|j_1\pm 1\rangle$ and $|j\pm 2\rangle$ due to the nonlinearity effect. For instance, if one fixed the first oscillator state at $|0\rangle$ as a pure state, the final state of this oscillator will be found in the superposition state of $|1\rangle$ and $|2\rangle$, meaning that $H_N$ causes the final states to be mixed. In other words, the state $|0\rangle$ is mixed with $|1\rangle$ and $|2\rangle$.

In addition, it is necessary to calculate the energy of the oscillators using $E_j = \langle j_i|H_0+H_N|j_i\rangle$ [20], where $j_i = 1,2$. The oscillators associated energies due to the total Hamiltonian are given by:

$$E_{j_1} = \langle j_1|H_0 + H_N|j_1\rangle = \hbar\omega_1\left(j_1 + \frac{1}{2}\right) + \hbar\{2g_{N11}\text{Re}(A_2) - i2g_{N61}\text{Re}(A_2)\}$$

$$E_{j_2} = \langle j_2|H_0 + H_N|j_2\rangle = \hbar\omega_2\left(j_2 + \frac{1}{2}\right) + i\hbar\{2g_{N41}(2j_2+1)\}$$

(8)

From Eq. 8, it is clear that the first term is arisen due to the linear Hamiltonian, and the terms inside the curly bracket are generated because of the nonlinear Hamiltonian effect. It is called the perturbation effect on the energy levels of the oscillators. In this equation, $A_1$ and $A_2$ are the steady-state points (DC points) of $LC_1$ and $LC_2$, where the oscillators are designed to operate on. The DC points can be calculated using Heisenberg-Langevin equations in the steady-state [15, 16]; these points are essentially affected by the circuit's DC bias and the linear Hamiltonian effect. The DC points related to the circuit are calculated and presented in Appendix A (Eq. A7). It is clear from the mentioned equation that the circuit's steady state points strongly depend on the circuit's generated noise (shown in Fig. 1), and the DC bias point. As an essential result of this research, it is found that the quantities such as $g_{12}'$, $g_{22}'$, and $C_{q1q2}$ handling the displacement in the circuit (linear Hamiltonian) strongly affect the steady-state points. This gives any engineer a degree of freedom to control and manipulate the steady-state point in which the circuit is established to be operated.

*Two-mode squeezed thermal state*

A squeezed-coherent state is created via the acting of the squeezed and displacement operators on the vacuum state defined as $|\alpha,\zeta\rangle = D(\alpha)S(\zeta)|0\rangle$, where $|0\rangle$ is the vacuum state [20, 25]. In this study, it can be easily shown that the coherent state can be generated by the linear part of the total Hamiltonian. In contrast, a squeezed state for each oscillator needs quadratic terms such as $a_i^2$ and $a_i^{+2}$ in the exponent. Nonetheless, for two-mode squeezing, the Hamiltonian should have the terms like $\{a_1a_2 + a_1^+a_2^+\}$ [20-25]. Therefore, the two-mode squeezed state becomes analyzed by the evolution of $\exp[H_Nt/i\hbar]$, where $H_N$ is defined in Eq. 4. Based on this definition, any quadratic terms like $\{a_1a_2 + a_1^+a_2^+\}$ in the Hamiltonian may generate two-mode squeezing. Consequently, using $\exp[H_Nt/i\hbar]$, the two-mode squeezing state of the nonlinear circuit studied is presented by:

$$S(\zeta_{12}) = \exp\left[-\hbar g_{N11}(a_2 + a_2^+)(a_1 - a_1^+)^2 + i\hbar g_{N41}(a_1 - a_1^+)(a_2 + a_2^+)^2\right.$$
$$\left. - i\hbar g_{N61}(a_1 - a_1^+)(a_1 + a_1^+)(a_2 + a_2^+)\right]\frac{t}{i\hbar} \quad (9)$$
$$S(\zeta_{12}) = \exp \zeta_{12}(a_2a_1 - a_2^+a_1^+)t$$

Eq. 9 shows that the coupled oscillator can generate two-mode squeezing through the transistor's nonlinearity. That means that the nonlinearity created by the transistor couples two oscillators so that the coupled two-mode become squeezed. The two-mode squeezing parameter is defined as $\zeta_{12} = -2g_{N11}\text{Im}(A_1) + 2g_{N41}\text{Re}(A_2) + 2g_{N61}\text{Re}(A_1)$, where Re{} and Im{} indicate the real and imaginary parts, respectively. It is apparent from the relationship that $\zeta_{12}$ can be a complex number, which means that the two-mode squeezing parameter contains a phase, which determines the angle of the quadrature.

It should be noted that "$t$" in the exponent ($\exp[Ht/j\hbar]$) can be determined from $t < \min\{1/\kappa_1, 1/\kappa_2\}$, where $\kappa_1$ and $\kappa_2$ are the first and second oscillators' decay rates. By selecting "$t$", the system is forced to generate squeezing before the resonator decaying [25].

*Quantum discord*

The calculation of the quantum discord initiates using two generalizations of the classical mutual information [1-7]. The mutual information is used primarily to evaluate the total correlations between subsystems [15]. In the first generalization, the quantum mutual information for two systems, A and B, is defined as $I(\rho_{AB}) = S(\rho_A) + S(\rho_B) - S(\rho_{AB})$, where $S(\rho_A) = -\text{Tr}(\rho_A\log_2\rho_A)$ is the von Neumann entropy of system A, and $S(\rho_{AB})$ is the conditional von Neumann entropy [5-7]. In the following, for the system studied in this work, the first oscillator is shortly called A, and the second one is called B. The conditional entropy arises because the measurement process disturbs the state on which a physical system is set. In other words, the applied measurement on subsystem B may change the state of subsystem A.

The second generalization introduces the entropic quantity C(A|B), by which the classical correlation in the joint state $\rho_{AB}$ is calculated. The classical correlation defines the maximum information about one subsystem depending on the measurement types applied to the other subsystem. The entropic quantity, by considering the generalization, is defined as $C(\rho_{AB}) = S(\rho_A) - S_{\min}(\rho_{AB})$, where the only difference between mutual quantum information is $S_{\min}(\rho_{AB})$. This term is the conditional minimized entropy of system A over

all possible measurements on system B. That is generally described as positive operator valued measures (POVMs) [5-7]. Thus, quantum discord is usually defined as $D(\rho_{AB}) = I(\rho_{AB}) - C(\rho_{AB})$, and by substituting from the above definitions, it becomes $D(\rho_{AB}) = S(\rho_B) - S(\rho_{AB}) + S_{min}(A|B)$.

Fortunately, a compact formula can be presented for two-mode squeezed thermal states (zero-mean Gaussian states) to reduce the covariance matrix (CM) into the standard form [5]. Consequently, the CM of the selected modes in the system can be presented in the form of the following matrix:

$$V_{AB} = \begin{pmatrix} (\tau b + \eta)I & \sqrt{\tau(b^2-1)}C \\ \sqrt{\tau(b^2-1)}C & bI \end{pmatrix} \qquad (10)$$

where $\mathbf{I} \equiv \text{diag}(1,1)$, $\mathbf{C} \equiv \text{diag}(1,-1)$, $a = n_{o1} + 0.5$, $b = n_{o2} + 0.5$, $\tau = d_{o12}^2/(b^2-1)$, and $\eta = a - (b_2* d_{o12}^2/(b^2-1))$ [9]. In these equations, a and b are the expectation value of the I/Q signals for two oscillators derived as $a \equiv <I_1(\omega)I_1(\omega)> = <Q_1(\omega)Q_1(\omega)>$, $b \equiv <I_2(\omega)I_2(\omega)> = <Q_2(\omega)Q_2(\omega)>$, where $I = (\delta a_j^+ + \delta a_j)/\sqrt{2}$ and $Q = (\delta a_j - \delta a_j^+)/i\sqrt{2}$, for subscript $j = 1,2$. In the listed relationships, $n_{o1}$, $n_{o2}$, and $d_{o12}$ are the output mean photon numbers of the first and second oscillator and the cross-correlation phase-sensitive, respectively. One can use the input-output formula [20] to calculate the output mean photon numbers in the form of $n_{o1} = 2\kappa_1 <\delta a_1^+\delta a_1> + <\delta a_{in-1}^+\delta a_{in-1}>$, $n_{o2} = 2\kappa_2 <\delta a_2^+\delta a_2> + <\delta a_{in-2}^+\delta a_{in-2}>$, $d_{o12} = 2\sqrt{(\kappa_1\kappa_2)} <\delta a_1\delta a_2>$. Finally, using Eq. 6, the mean photon numbers of the oscillators are calculated as $n_1 \equiv <\delta a_1^+\delta a_1>$, $n_2 \equiv <\delta a_2^+\delta a_2>$, and $d_{12} \equiv <\delta a_1\delta a_2>$ [9, 17], where $n_1$, $n_2$, and $d_{12}$ are, respectively, the first-, second oscillator mean photon number, the cross-correlation between two mentioned oscillators. It is supposed that $d_{12}$ is real-valued.

The output entropy associated with the heterodyne detection is equal to the average entropy of the output ensemble A [5]. Since entropy is invariant under displacements, it may write $S(A|het_B) = h(\tau + \eta)$, where $h(x) \equiv (x+0.5)\log_2(x+0.5) - (x-0.5)\log_2(x-0.5)$. In other words, the von Neumann entropy of an n-mode Gaussian state with CM expressed in Eq. 10 is calculated as $S(V_{AB}) = \sum_{i=1}^{N} f(v_i)$, where $v_i$ are the related Symplectic eigenvalues [6]. However, there is a heterodyne detection for which it is optimal for minimization of the output entropy, so the Gaussian discord is optimal. The "Gaussian quantum discord" of a two-mode Gaussian state $\rho_{AB}$, assuming a two-mode squeezed thermal state in this study, can be defined as the quantum discord satisfying the conditional entropy restricted to the generalized Gaussian POVMs on B [6].

Finally, the compact form of quantum discord, classical correlation, and quantum mutual information are given, respectively, by $D(\rho_{AB}) = h(b) - h(v_-) - h(v_+) + h(\tau + \eta)$, $C(\rho_{AB}) = h(a) - h(\tau + \eta)$, and $I(\rho_{AB}) = h(a) - h(v_-) - h(v_+)$, where $v_\pm$ is the Symplectic eigenvalue of the CM. The Symplectic eigenvalues are defined as $v_\pm = [\Delta \pm \sqrt{(\Delta^2-4D)}]/2$, where $\Delta = \det(a\mathbf{I}) + \det(b\mathbf{I}) + 2\det(d_{o12}\mathbf{I})$ and $D = \det(V_{AB})$; in the recent formulas det{:} stands for the matrix determinant. In the compact form of the equation defined for the quantum discord, the first term stands for the von Neumann entropy of the second oscillator in the system. The second and third terms define the von Neumann conditional entropy of the system. The last term in the equation is the effect of the classical correlation depending on the type of measurement performed on the second oscillator. As a result, in the system defined, the second oscillator entropy and the off-diagonal elements in the CM significantly affect the system's quantum discord. For this reason, in this article, we specially pay attention to the interferences between the oscillators and determine which quantities can affect the quantum discord. Perhaps the other critical factor that may be considered to enhance the quantum discord is h(τ + η), by which the classical correlation is decreased. As mentioned in [5-7], this critical factor strongly depends on the type of measurements.

**Results and Discussions**

In the following, the quantum correlation generated between the two-mode squeezing state by the circuit shown in Fig. 1 is discussed. It attempted to focus on the crucial parameters in the circuit (Fig. 2) that one can manipulate in such a way as to enhance the quantum correlation between modes. As shown in Fig. 2, $C_N$ and $g_{N2}$ are two crucial factors, mainly the function of $g_{m2}$ and $g_{m3}$, that the study concentrates on them to manipulate the quantum correlation. Notably, it was theoretically shown in the latter section that $g_{N2}$ is the main factor by which the final states of the oscillators became mixed and also plays a significant role in generating the two-mode squeezing thermal state.

Using the information mentioned above, in the following, we attempt to show that the quantum correlation can be generated between the microwave photons created by the two coupled oscillators. As a quantifier, the quantum discord is calculated, a property almost all quantum states hold. However, other quantities, such as the quantum information and classical correlation generated between modes in the system, are analyzed. The author thinks that comparing the quantum discord, quantum information, and classical correlation gives a solid sense for anyone to find which parameter can strongly enhance the quantum correlation. In addition, it makes it possible to know what portion of the classical correlation may appear in the quantum correlation. This means that all classically correlated states don't retain signatures of quantumness.

Before discussing the simulation results, it is necessary to care about a more severe and crucial point, which is InP HEMT operating at 4.2 K; this makes an astonishing thermally exciting noise in the circuit. Also, the internal circuit effects cause the quantum discord amplitude to remain limited. Thus, it seems impossible to compare the results of this study with the quantum discord generated by a typical opto-mechanical-microwave system's quantum discord [9]. It should be noted that the recent system works at 10 mK, whereas InP HEMT operates at 4.2 K; this means that the thermally excited photons dramatically affect the quantum correlation.

In this study, one of the aims is to design a circuit with an operating frequency ($f_{inc} = f_1+f_2$) of around 12 GHz, where $f_{inc}$, $f_1$, and $f_2$ are the RF incident frequency, the first and second oscillator's frequency, respectively. In the range of mentioned frequency and operating at 4.2 K, the mean thermally exciting photons were calculated as $n_{th1} \equiv <\delta a_{in\_1}^+ \delta a_{in\_1}>$ or $n_{th2} \equiv <\delta a_{in\_2}^+ \delta a_{in\_2}> \sim 12$, which is so greater than the average number of photons generated by the first and second oscillators $n_1 \equiv <\delta a_1^+ \delta a_1>$ or $n_2 \equiv <\delta a_2^+ \delta a_2>$ $\sim 0.2$. With knowledge of the thermally excited photons as the inserting noise and also the oscillators' mean photon number, in the following, the Gaussian quantum discord of the two-modes Gaussian states generated by the coupled oscillators is calculated. All simulations in this work containing the theoretical simulation of quantum correlation and also the simulation of the nonlinearity effects on the InP HEMT DC characteristics in PySpice were performed using the data from Table. 1.

Table. 1 Values for the small signal model of the 4*50 μm InP HEMT at 5 K [26-27].

|  | Stands for | Value [Unit] |
|---|---|---|
| $R_g$ | Gate resistance | 0.3 Ω |
| $L_g$ | Gate inductance | 75 pH |
| $L_d$ | Drain inductance | 70 pH |
| $C_{gs}$ | Gate-Source capacitance | 107 fF |
| $C_{ds}$ | Drain-Source capacitance | 51 fF |
| $C_{gd}$ | Gate-Drain capacitance | 60 fF |
| $R_i$ | Gate-Source resistance | 0.07 Ω |
| $R_j$ | Gate-Drain resistance | 8 Ω |
| $g_d$ | Drain-Source conductance | 12 mS |
| $g_m$ | Intrinsic transconductance | 82 mS |
| $V_g$ | Gate bias voltage | 0.03 V |
| $V_d$ | Drain bias voltage | 0.06 V |
| T | Operational temperature | 4.2 K |
| $T_d$ | Drain conductance noise temperature | 450 K |

The simulation results for quantum discord, the smaller Symplectic eigenvalue, quantum mutual information, and classical correlation at 4.2 K are shown in Fig. 3. The result in Fig. 3a indicates that increasing nonlinearity factor, $g_{N2}$, significantly enhances the quantum correlation. The graph shows that the quantum discord is strongly improved in the two different frequency ranges. It means that there is an avoided-level crossing in the graph (like a gap), where the quantum correlation between modes vanishes. According to the demonstrated figure, in some regions, $D(\rho_{AB})$ fluctuates around zero. It has been shown that a state with zero quantum discord represents a classical probability, while a positive discord indicates quantumness even in a separable mixed state [6, 7]. Therefore, for $0<D(\rho_{AB})<1$, the states may be entangled or separable, but for $D(\rho_{AB}) >1$, the states are entangled.

Generally, using the results illustrated, this paper wants to answer a few important questions: 1. Is it possible to generate quantum correlation using InP HEMT nonlinearity operating at 4.2 K? 2. By what factor the nonlinearity generated by the InP HEMT can affect the quantum correlation? 3. Is it possible to generate entangled microwave photons at 4.2 K? The Gaussian quantum A discord is calculated to answer the mentioned questions, and the measurement is performed on the second oscillator to minimize the classical correlation.

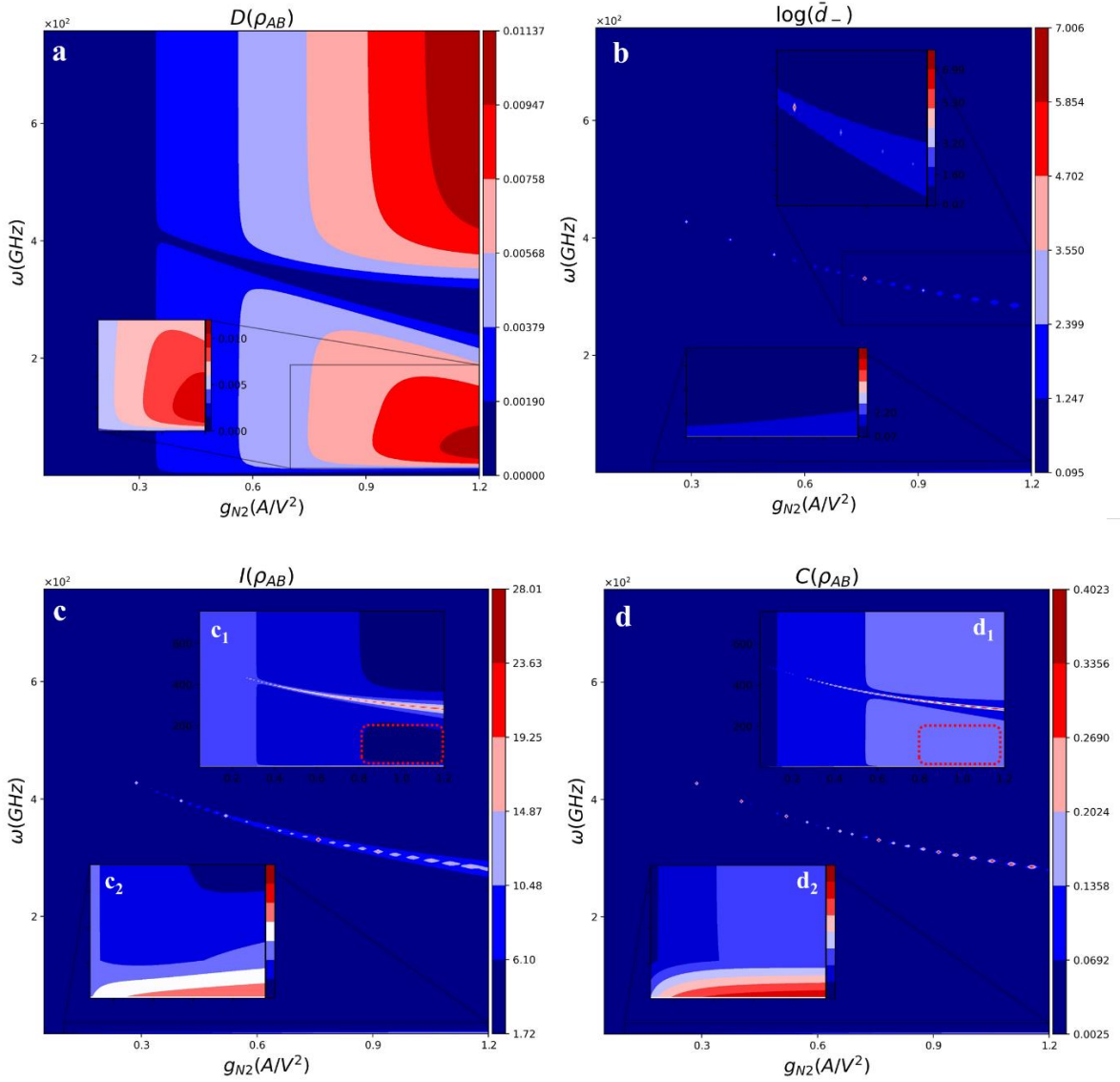

Fig. 3a) quantum discord, b) smaller Symplectic eigenvalue ($\log_{10}[\bar{d}_-]$), c) Quantum mutual Information; $c_1$ and $c_2$: $\log_{10}[I(\rho_{AB})]$, d) classical correlation; $d_1$ and $d_2$: $\log_{10}[C(\rho_{AB})]$, vs. angular frequency (GHz), and nonlinear factor $g_{N2}$ (A/V$^2$).

Fig. 3a demonstrates that the amplitude of quantum A discord is less than unity $0 < D(\rho_{AB}) < 1$, which means that the states generated by the oscillators may have either separable or entangled states. Consequently, if one applies a real InP HEMT operating at 4.2 K and considers all the noises and nonlinearity effects (Fig. 2), the generation of the entangled microwave photons seems impossible. This is mainly because more thermally excited photons are generated in microwave frequencies. Nonetheless, it is possible to create a partial quantum correlation between the two-mode Gaussian states (mixed states in this study). An inset figure is attached to Fig. 3a to show the critical points where the quantum discord is maximized. However, the interesting fact about this graph is the gap at which the quantum discord is minimized or completely disappeared. To discuss this odd behavior, the smaller Symplectic eigenvalue ($đ_-$) of the partially transposed state is illustrated in Fig. 3b. For better illustration, the logarithmic scale of the quantity ($\log_{10}(đ_-)$) is shown, at which two inset figures are attached. The minimum value for $đ_-$ is around 1.1 ($\log_{10}(1.1) = 0.095$). It has been shown that the Gaussian state is entangled if and only if $đ_- < 0.5$ [7] ($đ_- < 1$ [6]). As a result, the curve illustrated in Fig. 3b shows that the state generated by the circuit is not entangled because $đ_-$ is always greater than 1.1 for each $g_{N2}$. In addition, the study focuses on the smaller Symplectic eigenvalue because there is a pure consistency between the two figures illustrated in Fig. 3a and 3b. It can be found that the more quantum discord the circuit can generate, the less $đ_-$ the circuit can produce.

However, we learned from the latter section that some other quantifiers could help to elucidate the quantum discord's behavior. In other words, the difference between quantum discord, quantum mutual information, and classical correlation was mathematically discussed in the latter section. Thus, to thoroughly analyze the quantum discord, the simulation results of the quantum information and classical correlation are illustrated in Fig. 3c and Fig. 3d. The comparison between the quantum mutual information and classical correlation reveals that at the gap, where the quantum discord is minimized, the quantum mutual information and classical correlation are maximized. This means that at the gap mentioned, the modes totally become separable, and there is no quantumness between them. To discuss more technically, one may consider the definition of the Gaussian quantum discord, one-way classical correlation, and quantum mutual information as $D(\rho_{AB}) = h(b) - h(v_-) - h(v_+) + h(\tau + \eta)$, $C(\rho_{AB}) = h(a) - h(\tau + \eta)$, and $I(\rho_{AB}) = h(a) - h(v_-) - h(v_+)$ [5-7]. The common point between $D(\rho_{AB})$ and $I(\rho_{AB})$ is $-[h(v_-)+h(v_+)]$, which relates to the von Neumann entropy of the CM; however, the difference between them contributes to the oscillator's von Neumann entropy ($h(b)$ and $h(a)$), and $h(\tau + \eta)$ which just appears in $D(\rho_{AB})$. In Fact, the term $h(\tau + \eta)$ comes from the classical correlation effect, which minimizes the classical correlation to maximize the quantum correlation. Using the mentioned points and comparing Fig. 3a with Fig. 3c, one can find that the quantum discord is maximized where the quantum mutual information is minimized. The logarithmic inset graphs in Fig. 3c (Fig. 3c$_1$ and Fig. 3c$_2$) shows clearly the points.

The main difference between the $D(\rho_{AB})$ and $C(\rho_{AB})$ is the last term in the expressions $h(\tau + \eta)$, denoting the measurement effects on the second oscillators. The classical correlation is the maximum information about one subsystem, which depends on the measurement performed on the other subsystem. The result of the classical correlation is demonstrated in Fig. 3d. Also, some inset figures in the logarithmic scale are attached to clearly demonstrate the classical correlation as the function of circuit nonlinearity and frequency. This figure also shows that the classical correlation is maximized at the gap where the quantum discord disappeared. This effect contributes to the influence of the first oscillator's von Neumann entropy. The same effect could be found in the quantum mutual information graph.

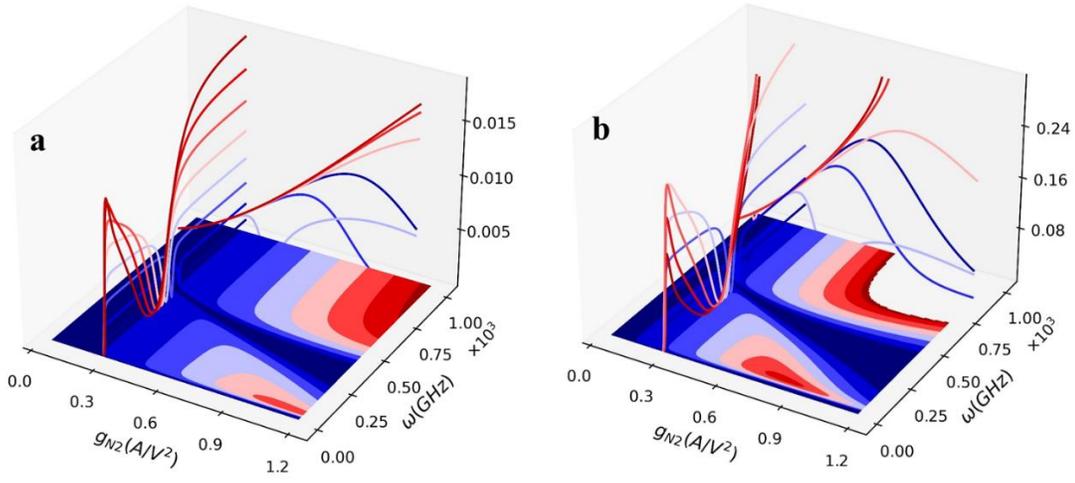

Fig. 4 quantum discord vs. angular frequency (GHz) and nonlinear factor $g_{N2}$ (A/V$^2$). a) $C_2 = 0.5$ pF, T = 4.2 K, b) T = 1.2 K; $C_2 = 0.5$ pF.

It mentioned that this study calculates the Gaussian quantum A discord, meaning that the measurement is performed on the second oscillator. As a result, if one wants to enhance the quantum A discord, it is necessary to minimize the thermally excited photons in the second oscillator. Thus, if one puts most of the thermal photons on the first oscillator (unmeasured subsystem) and decreases the thermal photons in the second oscillator, this enhances the Gaussian quantum discord. In the latter design, $C_2 = 1.0$ pF, by which the first and second oscillator's mean thermally excited photons for a typical frequency were, respectively, $n_{th1} = 11.74$, $n_{th2} = 13.01$; also the mean photon number (phase cross-correlation) for two coupled oscillators was around $n_{12} \equiv <\delta a_1 \delta a_2> = 2.63e-4$. Since $C_2$ is changed to 0.5 pF, the mean number of the thermally excited photons in the same frequency becomes $n_{th1} = 11.73$, $n_{th2} = 9.38$, and $n_{12} = 9.95e-4$. The estimated results show that the quantum A discord should be increased. The simulation result is shown in Fig. 4a as a 3D graph. It is shown that the quantum discord amplitude is increased. This is because the second oscillator's thermally excited photons are decreased. Using illustrated 3D figure, one can get more information about the Gaussian quantum discord since the change of the quantifier as the 1D curves in the x-axis and y-axis are annexed in the figures. The other graph in Fig. 4b shows the operating temperature effect on the quantum A discord. We suppose that if InP HEMT could operate at 1.2 K, what happens on the quantum A discord? Fig. 3b shows a significant enhancement in quantum discord amplitude. This directly contributes to the level of the thermally excited noise generations, which dramatically decreases as the temperature drop from 4.2 K to 1.2 K. The other interesting point is that by reducing the temperature to around 1.2 K, at lower $g_{N2}$ the quantum discord amplitude is increased. In the following, we will study that increasing $g_{N2}$ can dramatically change the InP HEMT DC characterization, by which, for instance, the power dissipation can be severely increased. In other words, manipulating $g_{N2}$ to enhance the quantum discord adds a crucial trade-off to RF circuit engineering.

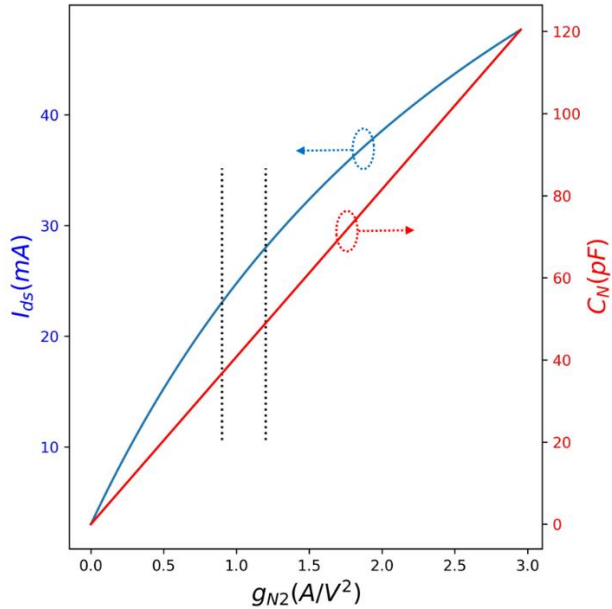

Fig. 5 InP HEMT DC characterization: $I_{ds}$ (mA) vs. $g_{N2}$ (A/V$^2$), Nonlinear capacitance $C_N$ (pF) vs. $g_{N2}$ (A/V$^2$)

So far, the study shows that by engineering some parameters related to the internal circuit of the InP HEMT specially focused on the nonlinearity factors, it was possible to enhance the quantum correlation between the two-mode Gaussian states. Initially, it was theoretically shown that the states generated by the InP HEMT nonlinear circuit became mixed. The results showed that the quantum discord is increased for $g_{N2}$ between 0.8-1.2 A/V$^2$, meaning that the quantum correlation may be created between the states. Finally, arising a trade-off in the design was discussed, which means that one cannot increase $g_{N2}$ to any value to get the desired quantum correlation. It is because the nonlinearity factors can dramatically change the modified circuit performance (DC and AC characteristics). In other words, by changing the nonlinearity factors such as $g_{m2}$ and $g_{m3}$, for example, the DC characteristics of the InP HEMT can be changed. The mentioned variations contribute to the circuit's components changing, such as $C_N$ and $g_{N2}$. To show this point, the circuit illustrated in Fig. 2 is simulated in PySpice (Jupyter Lab) using the data from Table.1. We simulated the internal circuit of InP HEMT, and the related DC characterization is shown in Fig. 5. The left curve as a blue graph shows the InP HEMT drain-source current ($I_{ds}$) as the function of $g_{N2}$, whereas the right graph (red) displays the $C_N$ as the function of $g_{N2}$. In the region shown between the dotted line, where the $g_{N2}$ is changed between 0.8-1.2, $I_{ds}$ is increased to ~30 mA, and $C_N$ is reached around 50 pF. It is better to note that $I_{ds}$ is altered in the range of (0.5- 2.3) mA for sub-mW cryogenic applications [26]. The mentioned trade-off becomes crucially created, and the designers should especially care about this point. In other words, increasing the current to 30 mA may generate the quantum correlation between states. Still, it forces the system to dissipate energy dramatically, which is a crucial point in cryogenic applications.

As an essential conclusion of this work, it can be mentioned that InP HEMT nonlinearity has the potential to partially generate the quantum correlation between modes of the oscillators coupled through the transistor. Some critical factors play a central role, such as the nonlinear capacitor arising due to the nonlinearity $C_N$ and the nonlinearity factor $g_{N2}$ affected by $g_{m2}$ and $g_{m3}$. These parameters give engineers a reasonable degree of freedom to effectively design a cryogenic circuit containing InP HEMT by which the generation of the quantum correlation between modes at 4.2 becomes possible. However, the generation of entangled microwave photons by InP HEMT operating at 4.2 K seems impossible, at least with the recent technology.

**Conclusions**

This study investigated the quantum correlation of the microwave two-mode squeezed thermal state generated by the nonlinearity of InP HEMT. For this reason, the nonlinear circuit related to InP HEMT was analyzed using quantum theory, and the contributed dynamics equation of the motion of the circuit was theoretically derived. In this work, we intensely concentrated on the nonlinear Hamiltonian by which it was theoretically shown that the two-mode squeezed thermal state was generated. In addition, for the circuit discussed, it was theoretically proved that the circuit just generated the mixed state, and there was no pure state. Because of the facts mentioned, the study focused on quantum discord rather than quantum entanglement. To completely know about the quantum discord, other quantities such as quantum mutual information, classical correlation, and the smaller Symplectic eigenvalue were analyzed.

Some engineering was carried out on the nonlinear circuit to get the desired results, and some critical parameters, such as $g_{m2}$ and $g_{m3}$, were manipulated to enhance the quantum correlation between the generated modes. As an important conclusion, the simulation result showed that although it might be possible to improve the quantum correlation between modes, attaining quantum discord greater than unity seems challenging when InP HEMT operated at 4.2 K.

Appendix A:

The linear Hamiltonian of the system is given by:

$$H_L = \frac{1}{2C_{q1}}Q_1^2 + \left(\frac{1}{2L_1}\right)\varphi_1^2 + \frac{1}{2C_{q2}}Q_2^2 + \left(\frac{1}{2L_2} + \frac{1}{2L_{p2}} - \frac{2g_m g_{N2} C_B^2 C_{in} V_{rf}}{C_M^4}\right)\varphi_2^2 + \frac{1}{C_{q1q2}}Q_1 Q_2$$

$$+ \left(g_{12} + \frac{2g_{N2} C_B^2 C_{in} V_{rf}}{C_M^4}\right)Q_1\varphi_2 + \left(g_{22} + \frac{2g_{N2} C_B C_c C_{in} V_{rf}}{C_M^4}\right)Q_2\varphi_2 - \overline{I_{gs}^2}\varphi_1 + V_{q1}Q_1 + V_{q2}Q_2$$

$$+ \left(I_{P2} + \frac{2g_{N2} C_B^2 C_{in}^2 V_{rf}^2}{C_M^4}\right)\varphi_2 - \frac{C_{gs}}{2}\overline{V_i^2};$$

(A1)

$$\frac{1}{L_{2'}} \equiv \frac{1}{2L_2} + \frac{1}{2L_{p2}} - \frac{2g_m g_{N2} C_B^2 C_{in} V_{rf}}{C_M^4}, \quad g_{12}' \equiv g_{12} + \frac{2g_{N2} C_B^2 C_{in} V_{rf}}{C_M^4}, \quad g_{22}' \equiv g_{22} + \frac{2g_{N2} C_B C_c C_{in} V_{rf}}{C_M^4}$$

where $C_{q1}$, $C_{q2}$, $C_{q1q2}$, $L_{p2}$, $g_{12}$, $g_{22}$, $V_{q1}$, $V_{q2}$, and $I_{p2}$ are constants and are defined as:

$$C_M^4 = C_B(C_{A'} + C_N) - C_c^2; \quad C_{A'} = C_A + C_N; \quad C_N = g_{m2}\varphi_{2dc} + 6g_{m3}\varphi_{2dc}\dot{\varphi}_{1dc}$$

$$\overline{I_{gs}^2} = \overline{I_g^2} - \overline{I_j^2}; \quad \overline{I_{ds}^2} = \overline{i_{ds}^2} + \overline{I_d^2} + \overline{I_j^2}$$

$$\frac{1}{C_{q1}} = \frac{1}{C_M^4}\{C_B^2 C_A - C_c^2 C_B\}, \quad \frac{1}{C_{q2}} = \frac{1}{C_M^4}\{C_c^2 C_A + C_{A'}^2 C_B - 2C_c^2 C_{A'}\}$$

$$\frac{1}{C_{q1q2}} = \frac{1}{C_M^4}\{C_B C_c C_A - C_c^3\}, \quad \frac{1}{L_{p2}} = \frac{1}{C_M^4}\{g_m^2 C_B^2 C_A - 2g_m^2 C_c^2 C_B\}$$

(A2)

$$g_{12} = \frac{1}{2C_M^4}\{-2g_m C_B^2 C_A + 3g_m C_c^2 C_B\}, \quad g_{22} = \frac{1}{C_M^4}\{-g_m C_B C_c C_A + g_m C_B C_c C_{A'} + g_m C_c^3\}$$

$$V_{q1} = \frac{1}{C_M^4}\{C_B^2 C_{in} C_A V_{rf} - C_B C_{in} C_c^2 V_{rf}\}, \quad V_{q2} = \frac{1}{C_M^4}\{C_B C_c C_{in} C_A V_{rf} + 0.5 C_B C_c C_{in} C_{A'} V_{rf} - C_{in} C_c^3 V_{rf}\}$$

$$I_{p2} = \frac{1}{C_M^4}\{-g_m C_B^2 C_{in} C_A V_{rf} + g_m C_B C_c^2 C_{in} V_{rf}\} - \overline{I_{ds}^2}$$

$$\begin{cases} g_{N11} = \frac{g_{N2}}{C_M^4}\frac{C_B^2}{2Z_1}\sqrt{\frac{\hbar Z_2}{2}}, & g_{N21} = \frac{g_{N2}}{C_M^4}\frac{C_c^2}{2Z_2}\sqrt{\frac{\hbar Z_2}{2}} \\ g_{N31} = \frac{g_{N2} g_m^2}{C_M^4}\frac{Z_2 C_B^2}{2}\sqrt{\frac{\hbar Z_2}{2}}, & g_{N41} = \frac{g_{N2} g_m}{C_M^4}\frac{Z_2 C_B^2}{2}\sqrt{\frac{\hbar}{2Z_2}} \\ g_{N51} = \frac{g_{N2} g_m}{C_M^4}\frac{Z_2 C_B C_c}{2}\sqrt{\frac{\hbar}{2Z_2}}, & g_{N61} = \frac{g_{N2} g_m}{C_M^4}\frac{\sqrt{Z_1 Z_2} C_B C_c}{2}\sqrt{\frac{\hbar}{2Z_1}} \end{cases}$$

(A3)

$$\gamma_{a11} = -4ig_{N11}\,\text{Re}\{A_2\}$$

$$\gamma_{a12} = -4ig_{N11}\,\text{Re}\{A_1\} - 4g_{N41}\,\text{Re}\{A_2\} + 2g_{N61}\{A_1\}^*$$

$$\gamma_{a13} = 4g_{N61}\,\text{Re}\{A_2\}$$

$$\gamma_{a21} = -4g_{N11}\,\text{Im}\{A_2\} + 2ig_{N21}\,\text{Re}\{A_2\} + 4g_{N41}\,\text{Re}\{A_2\} - 2g_{N61}\,\text{Re}\{A_1\} \quad (A4)$$

$$\gamma_{a22} = -2g_{N21}\,\text{Im}\{A_1\} + 12ig_{N31}\,\text{Re}\{A_2\} + 4ig_{N41}\,\text{Im}\{A_1\} - 4g_{N51}\,\text{Re}\{A_2\} + 4ig_{N51}\,\text{Im}\{A_2\}$$

$$\gamma_{a22} = 4g_{N51}\,\text{Re}\{A_2\}$$

$$\gamma_{a23} = -2ig_{N61}\,\text{Im}\{A_1\}$$

$$j_{p21} = \hbar(g_{N21} + g_{N31} + ig_{N51}),\; j_{p22} = \hbar(g_{N21} + g_{N31} - ig_{N51})$$

$$j_{p23} = \hbar\left(-g_{N21} + 4g_{N61}\{A_1\}^* + ig_{N51}\right),\; j_{p24} = \hbar\left(-g_{N21} + 4g_{N61}\{A_1\}^* - ig_{N51}\right)$$

$$j_{p25} = \hbar(g_{N21} + g_{N31} + ig_{N51}),\; j_{p26} = \hbar(g_{N21} + g_{N31} - ig_{N51}) \quad (A5)$$

$$j_{p27} = j_{p27} = \hbar g_{N31}$$

$$j_{p28} = \hbar(g_{N31} + 2ig_{N51}),\; j_{p210} = \hbar(g_{N31} - 2ig_{N51})$$

$$j_{p211} = j_{p212} = -2\hbar g_{N41}\,\text{Im}(A_1)$$

Using the Hamiltonian expressed in A1 and the related dynamic equation of motion, DC points (constant points) of the circuit can be calculated. Therefore, the steady state equations become:

$$0 = -\left(i\Delta_1 + \frac{\kappa_1}{2}\right)A_1 - \frac{i}{2C_{q1q2}\sqrt{Z_1 Z_2}}(A_2 - A_2^*) + \sqrt{\frac{Z_2}{Z_1}}g_{12}'(A_2 + A_2^*) + V_{q1}\sqrt{\frac{1}{2\hbar Z_1}} + i\overline{I_{gs}^2}\sqrt{\frac{Z_1}{2\hbar}}$$

$$0 = -\left(i\Delta_2 + \frac{\kappa_2}{2}\right)A_2 - \frac{i}{2C_{q1q2}\sqrt{Z_1 Z_2}}(A_1 - A_1^*) - \sqrt{\frac{Z_2}{Z_1}}g_{12}'(A_1 - A_1^*) + g_{22}'(A_2^*) + V_{q2}\sqrt{\frac{1}{2\hbar Z_2}} - i\overline{I_{p2}^{'2}}\sqrt{\frac{Z_2}{2\hbar}} \quad (A6)$$

Using the standard approach to solve the equations, the final form of the variables become:

$$\begin{bmatrix} -\dfrac{\kappa_1}{2} & \Delta_1 & 2g_{12}'\sqrt{\dfrac{Z_2}{Z_1}} & \dfrac{1}{2C_{q1q2}\sqrt{Z_1 Z_2}} \\ -\Delta_1 & \dfrac{\kappa_1}{2} & 0 & 0 \\ 0 & \dfrac{1}{2C_{q1q2}\sqrt{Z_1 Z_2}} & g_{22}' - \dfrac{\kappa_2}{2} & \Delta_2 \\ 0 & -2g_{12}'\sqrt{\dfrac{Z_2}{Z_1}} & -\Delta_2 & -\left(g_{22}' + \dfrac{\kappa_2}{2}\right) \end{bmatrix} \begin{bmatrix} A_{1R} \\ A_{1j} \\ A_{2R} \\ A_{2j} \end{bmatrix} = \begin{bmatrix} -V_{q1}\sqrt{\dfrac{1}{2\hbar Z_1}} \\ -\overline{I_{gs}^{\,2}}\sqrt{\dfrac{Z_1}{2\hbar}} \\ -V_{q2}\sqrt{\dfrac{1}{2\hbar Z_2}} \\ \overline{I_{p2}^{'2}}\sqrt{\dfrac{Z_2}{2\hbar}} \end{bmatrix} \quad (A7)$$

where $A_1 = A_{1R} + iA_{1j}$ and $A_2 = A_{2R} + iA_{2j}$. Using Eq. A7 the steady state point of the circuit, which is related to the thermal noise and bias point of the circuit, is calculated.